\documentclass[conference]{IEEEtran}
\IEEEoverridecommandlockouts
\usepackage{cite}
\usepackage{amsmath,amssymb,amsfonts}

\usepackage{graphicx}
\usepackage{textcomp}
\usepackage{xcolor}
\usepackage{cite}

\usepackage{fancyhdr}
\usepackage{comment}
\usepackage{multicol}
\usepackage{afterpage}
\usepackage{subcaption}
\usepackage{amsmath,amssymb,amsfonts}
\usepackage{algorithm}
\usepackage{algpseudocode}
\usepackage{graphicx}
\usepackage{textcomp}
\usepackage{orcidlink}
\usepackage{tabularx}
\usepackage{xcolor}
\def\BibTeX{{\rm B\kern-.05em{\sc i\kern-.025em b}\kern-.08em
    T\kern-.1667em\lower.7ex\hbox{E}\kern-.125emX}}
\begin{document}

\title{Enhancing Communication Efficiency in FL with Adaptive Gradient Quantization and Communication Frequency Optimization \\

\thanks{This work is supported by the Zayed health science center under fund 12R005. A.Tariq (Email: 700039114@uaeu.ac.ae), F.Sallabi (Email: F.sallabi@uaeu.ac.ae), T. Qayyum (Email: 700036923@uaeu.ac.ae) and E. Barka is with College of IT, United Arab Emirates University, UAE. M.A.Serhani (Email: mserhani@uaeu.ac.ae) is with College of Computing and Informatics, UoS. UAE. I.Taleb (Email: ikbal.taleb@zu.ac.ae) is with CTI, Zayed University, Abu Dhabi, UAE.}
\thanks{\textit{Corresponding Author: \textit{Ezedin S. Barka} (Email: ebarka@uaeu.ac.ae) and \textit{Mohamed Adel Serhani} (Email: mserhani@sharjah.ac.ae)}}
}

\author{
    \IEEEauthorblockN{Asadullah Tariq\IEEEauthorrefmark{1}, Tariq Qayyum\IEEEauthorrefmark{1}, Mohamed Adel Serhani\IEEEauthorrefmark{2}, Farag M. Sallabi\IEEEauthorrefmark{1}, Ikbal Taleb\IEEEauthorrefmark{3},
    Ezedin S. Barka\IEEEauthorrefmark{1}}
   
    \IEEEauthorblockA{\IEEEauthorrefmark{1}College of Information Technology, United Arab Emirates University, Al Ain, Abu Dhabi, UAE.\\
    }
    
    \IEEEauthorblockA{\IEEEauthorrefmark{2}College of Computing and Informatics, University of Sharjah, Sharjah, UAE.\\
    }

    \IEEEauthorblockA{\IEEEauthorrefmark{3}College of Technological Innovation, Zayed University, Abu Dhabi.UAE.\\
   }
}



\maketitle
\fancypagestyle{firstpageheader}{
    \fancyhf{} 
    \renewcommand{\headrulewidth}{0pt} 
    \fancyhead[C]{\sffamily \textcolor{blue}{Accepted at International Conference on Communications \& (ICC) 2025, Montreal, Canada.}}
}
\thispagestyle{firstpageheader}
\begin{abstract}
Federated Learning (FL) enables participant devices to collaboratively train deep learning models without sharing their data with the server or other devices, effectively addressing data privacy and computational concerns. however, FL faces a major bottleneck due to high communication overhead from frequent model updates between devices and the server, limiting deployment in resource-constrained wireless networks. In this paper, we propose a three-fold strategy: firstly, an Adaptive Feature-Elimination Strategy to drop less important features while retaining high-value ones; secondly, Adaptive Gradient Innovation and Error Sensitivity-Based Quantization, which dynamically adjusts the quantization level for innovative gradient compression; and thirdly, Communication Frequency Optimization to enhance communication efficiency. We evaluated our proposed model’s performance through extensive experiments, assessing accuracy, loss, and convergence compared to baseline techniques. The results show that our model achieves high communication efficiency in the framework while maintaining accuracy.
\end{abstract}


\section{Introduction and background}

In Federated Learning (FL), the server updates its learning parameters based on data insights (typically gradients) received from local devices (clients) and redistributes these updated parameters back to the clients for federation. This setup enables the server to gather aggregated insights without requiring direct access to raw data, thereby supporting privacy and lowering computational demands on the server \cite{b1}. However, this model introduces communication overhead that becomes increasingly inefficient as the number of participating clients grows. This problem becomes more challenging as the complexity of the global model on the server increases due to the increasing number of model parameters \cite{b2}.

Communication involves two primary phases in FL: distributing model parameters from the server to clients (downlink) and transmitting updated parameters from clients back to the server (uplink). The uplink transmission poses significant and much tighter communication bottleneck than downlink bandwidth. Downlink communication in FL is more bandwidth-efficient, as the same set of parameters can be broadcast to all clients simultaneously. In contrast, the uplink phase requires the server to collect diverse parameter updates from multiple clients, which demands greater bandwidth and coordination \cite{b3}.

Gradient compression is necessary to address these communication bottleneck problems. Various compression methods have been widely introduced, where these techniques commonly apply lossy compression to the local gradients of model parameters calculated on each device \cite{b4}. Two key methods for gradient compression are gradient quantization \cite{b5} and gradient sparsification \cite{b6}. The  gradient quantization methods in FL reduce transmission bit requirements by compressing gradients. However, sparsification schemes in FL only transmit a subset of gradients that notably influence parameter updates during each communication round. FedPAQ in \cite{gq1}, leverages periodic averaging, quantization of updates, and partial node participation. The Adaptive Quantized Gradient (AQG) method in \cite{gq2} adjusts quantization levels based on local gradient updates to minimize unnecessary transmissions and addresses client dropouts by introducing an augmented AQG. DAdaQuant in \cite{gq3} builds on the QSGD fixed-point quantizer, introducing a doubly-adaptive quantization approach that adjusts quantization levels both over time and across clients to enhance communication compression in FL. The authors in \cite{gq4} focus on gradient compression through block sparsification, dimensionality reduction, and quantization. The Lazily Aggregated Quantized (LAQ) method in \cite{gq5} selectively quantizes and reuses gradients, reducing the need for frequent updates. The work in \cite{gq6} presents a decentralized compressed variant to further enhance communication efficiency. FedDQ in \cite{mq4}, introduces adaptive quantization levels based on shrinking model updates during training, thus reducing communication volume and rounds. The research work in \cite{a1}, \cite{a2} proposed a data quality based dynamic data sample selection and client selection leveraging explainable AI and game nnash equilibrium. The adaptive quantization method in AdaGQ \cite{gq8} adjusts resolution based on gradient norm variations, while heterogeneous quantization assigns lower resolution to slower clients and higher resolution to faster ones. Quantization can also be used with sparsification, as in SOBAA \cite{gq9}, \cite{new}, which combines layer-wise sparsification, one-bit quantization, power control, and an error-feedback mechanism to improve convergence. These methods optimize the compression ratio and power control to minimize errors, accelerating convergence. 

Current gradient quantization algorithms typically use predetermined and fixed quantization levels during the training process. However, each FL task presents unique characteristics regarding communication costs, convergence time, and network conditions. This makes fixed quantization less effective, highlighting the need to develop more dynamic and adaptive quantization algorithms. Additionally, solutions should be implemented to better control the exchange of model gradients between clients and servers to enhance communication efficiency. To address the communication bottleneck problem discussed above, we propose a three-fold communication-efficient framework comprising an Adaptive Feature-Elimination Strategy, Adaptive Gradient Innovation and Error Sensitivity-Based Quantization, and Communication Frequency Optimization. The main contributions of our paper are as follows:

\begin{enumerate}
\item The Adaptive Feature-Elimination Strategy compresses feature vectors selectively by probabilistically dropping less informative vectors, thus retaining high-value features.

\item Adaptive Gradient Innovation and Error Sensitivity-Based Quantization compresses gradient updates by quantizing only the innovations (changes) in gradients for further trainings and dynamically adjusts the quantization level for each training round based on the error sensitivity.

\item The Communication Frequency Optimization further enhances the quantization strategy by allowing clients to selectively withhold updates that do not significantly contribute to model improvements.

\item The evaluation of our proposed model’s performance through extensive experiments, assessing accuracy, loss, and convergence compared to baseline techniques. 
\end{enumerate}

The remainder of the paper is organized as follows: Section 2 presents the main methodology, Section 4 covers the results and experimental evaluation, and Section 5 discusses the conclusion and future work.

\section{Proposed Methodology}

\begin{figure*}[!ht]
\centering
\includegraphics[width=15.3cm,height=9cm,keepaspectratio]{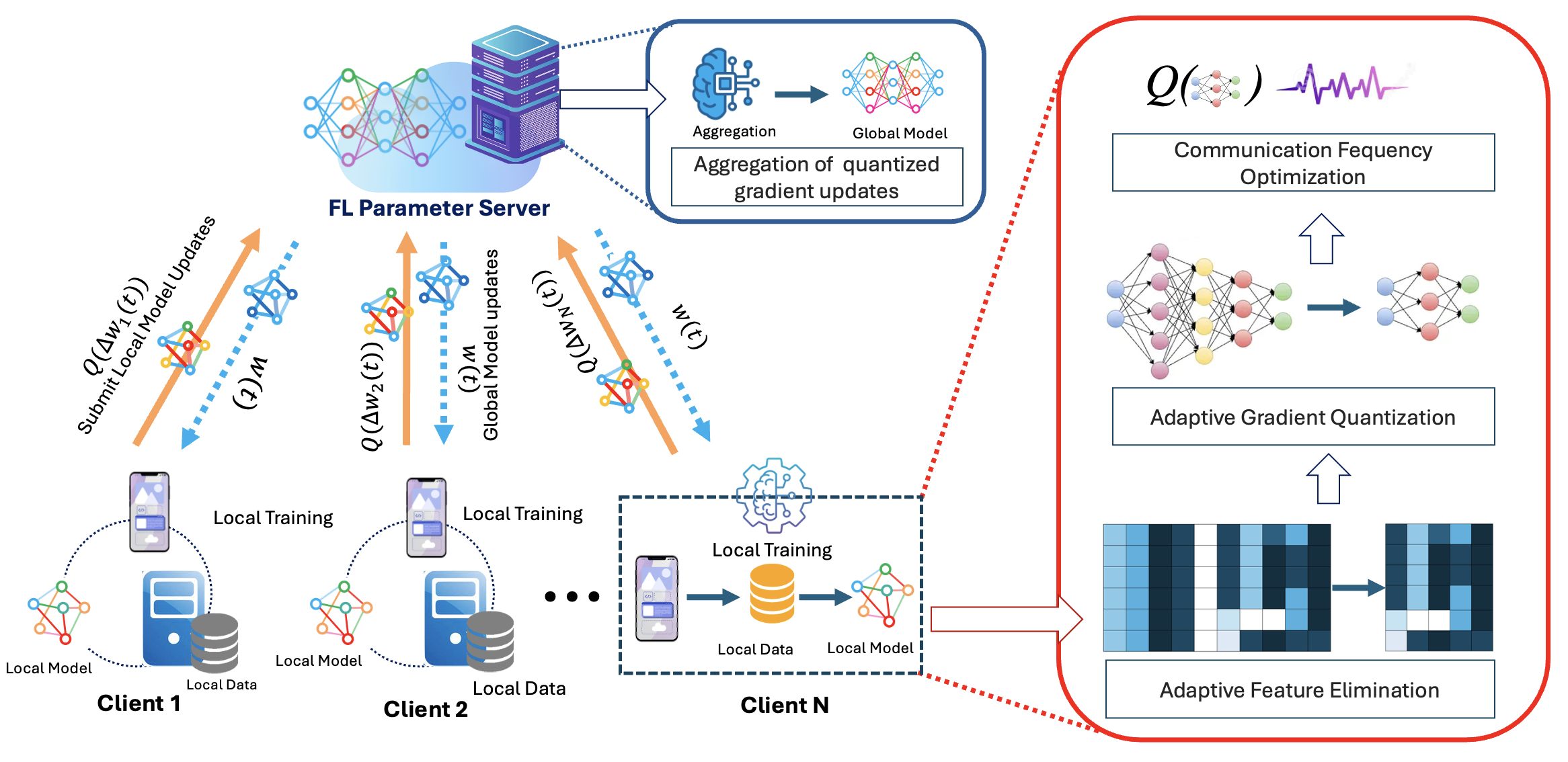}
\caption{Proposed framework presenting Communication Efficient Adaptive Gradient Quantization and Communication Frequency Optimization  }
\label{fig1}
\end{figure*}

In this section, we present our proposed three-fold methodology—a communication-efficient framework comprising an Adaptive Feature-Elimination Strategy, Adaptive Gradient Innovation and Error Sensitivity-Based Quantization, and Communication Frequency Optimization, as depicted in Fig.~\ref{fig1}.
\subsection{Adaptive Feature-Elimination}

The Adaptive Feature-Wise Dropout Strategy selectively compresses feature vectors by probabilistically dropping less informative ones. In each training round, each device \( m \in \{1, \ldots, M\} \) computes an intermediate feature matrix \( F^{(m)} \in \mathbb{R}^{B \times D} \). Here, \( B \) is the batch size, and \( D \) is the number of feature dimensions. Each column \( f_i^{(m)} \in \mathbb{R}^B \) of \( F^{(m)} \) represents the feature vector for the \( i \)-th feature across \( B \) samples in the batch.

To measure the importance of each feature vector \( f_i^{(m)} \), we compute its standard deviation across the batch:
\begin{equation}
\sigma_i^{(m)} = \sqrt{\frac{1}{B} \sum_{b=1}^B \left( f_{i,b}^{(m)} - \mu_i^{(m)} \right)^2}
\end{equation}

where \( \mu_i^{(m)} = \frac{1}{B} \sum_{b=1}^B f_{i,b}^{(m)} \) is the mean of the \( i \)-th feature vector. To normalize the importance of each feature vector, we calculate:
\begin{equation}
q_i^{(m)} = \frac{\sigma_i^{(m)} D}{\sum_{j=1}^D \sigma_j^{(m)}}
\end{equation}

The dropout probability \( p_i^{(m)} \) for each feature vector \( f_i^{(m)} \) is then:
\begin{equation}
p_i^{(m)} = \begin{cases} 
      1 - q_i^{(m)}, & \text{if } \max(q_j^{(m)}) \leq 1 \\
      1 - \frac{(\sigma_i^{(m)} + C_{\text{bias}}) D}{\sum_{j=1}^D (\sigma_j^{(m)} + C_{\text{bias}})}, & \text{if } \max(q_j^{(m)}) > 1 
   \end{cases}
\end{equation}

The bias term \( C_{\text{bias}} \) is set to ensure that dropout probabilities remain within \([0,1]\):
\begin{equation}
C_{\text{bias}} \geq \frac{\sigma_{\max}^{(m)} D - \sum_{j=1}^D \sigma_j^{(m)}}{D - D'}
\end{equation}

Using the dropout probabilities \( p_i^{(m)} \), each device generates a binary indicator \( \delta_i^{(m)} \) for each feature vector:
\begin{equation}
\delta_i^{(m)} \sim \text{Bernoulli}(1 - p_i^{(m)})
\end{equation}

The expected number of retained feature vectors, \( D^{(m)}_{\text{retained}} \), is:
\begin{equation}
D^{(m)}_{\text{retained}} = \sum_{i=1}^D (1 - p_i^{(m)}) = D - \sum_{i=1}^D p_i^{(m)}
\end{equation}

To maintain the expected value, each retained feature vector \( f_i^{(m)} \) is scaled by:
\begin{equation}
\hat{f}_i^{(m)} = \frac{\delta_i^{(m)}}{1 - p_i^{(m)}} f_i^{(m)}
\end{equation}

\subsection{Adaptive Gradient Innovation and Error Sensitivity-Based Quantization}

Gradient Innovation-Based Quantization focuses on compressing gradient updates by quantizing only the innovations (changes) in gradients from one training round to the next. The innovation for client \( m \) at iteration \( k \) is represented by the difference between the current and previous gradients, reducing the amount of redundant data transmitted when gradient changes are minimal.

Let \( w^{(k)} \) represent the model parameters at the \( k \)-th iteration, and \( \nabla f^{(k)}_m(w) \) be the gradient computed by client \( m \) based on its local data. The gradient innovation \( \Delta g_m^{(k)} \) for client \( m \) at iteration \( k \) is defined as the difference between the current gradient and the quantized gradient of the previous iteration:
\begin{equation}
\Delta g_m^{(k)} = \nabla f^{(k)}_m(w) - Q\left(\nabla f^{(k-1)}_m(w)\right)
\end{equation}
where \( Q(\cdot) \) is the quantization operator that reduces the precision of the gradient vector, allowing each element of \( \Delta g_m^{(k)} \) to be represented in a lower bit depth. The quantized gradient innovation allows us to represent only the essential changes in the gradient, thus conserving communication. To further reduce data volume, we apply a quantization function to the gradient innovation \( \Delta g_m^{(k)} \). Let \( b \) be the number of bits allocated to each quantized component of the gradient innovation vector. Each element \( \Delta g_{m,i}^{(k)} \) of the gradient innovation is quantized as follows:
\begin{equation}
q_{m,i}^{(k)} = \text{Quantize}\left(\Delta g_{m,i}^{(k)}; b\right) = \left\lfloor \frac{\Delta g_{m,i}^{(k)} + R}{2 \cdot t \cdot R} + \frac{1}{2} \right\rfloor
\end{equation}

This equation effectively scales \( \Delta g_{m,i}^{(k)} \) to fit within a specified range and compresses each entry into \( b \)-bit integers, leading to significant data savings. The quantization ensures that only a few bits are used to represent each component of the innovation vector.

However, a fixed quantization level can lead to inefficient resource use, especially in FL settings where the training dynamics can vary significantly. To enhance this, we incorporate the Error Sensitivity-Based Quantization scheme, which adapts the quantization level \( q_t \) over time based on a dynamically calculated error-sensitivity metric \( E_t \), reflecting the model’s current sensitivity to quantization errors. This model builds on the our adaptive quantization of \( \Delta g_m^{(k)} \) by dynamically adjusting the quantization level \( q_t \) for each training round \( t \). Rather than using a static quantization level, it modifies \( q_t \) based on the error sensitivity of the current training state, which helps maintain high accuracy even under aggressive compression. The error sensitivity \( E_t \) is computed as the variance of client gradients around the global model gradient. Define the global model gradient \( G_t \) at round \( t \) as:
\begin{equation}
G_t = \frac{1}{|S_t|} \sum_{i \in S_t} \frac{1}{|D_i|} \nabla F_i(\mathbf{w}_t)
\end{equation}
where \( S_t \) is the set of sampled clients, \( D_i \) is the local dataset for client \( i \), and \( \nabla F_i(\mathbf{w}_t) \) represents the local gradient for client \( i \) on the model \( \mathbf{w}_t \). The error sensitivity \( E_t \) is then defined as:
\begin{equation}
E_t = \frac{1}{|S_t|} \sum_{i \in S_t} \left\| \nabla F_i(\mathbf{w}_t) - G_t \right\|^2
\end{equation}
If \( E_t \) is high, this indicates that the model is sensitive to quantization errors and requires a lower quantization level (higher precision). Conversely, a low \( E_t \) suggests that the model can tolerate a higher quantization level (coarser updates).

The quantization level \( q_t \) is adjusted based on \( E_t \) relative to a threshold \( E_{\text{thresh}} \). Using a scaling factor \( \gamma \) (where \( \gamma > 1 \)), the update rule for \( q_t \) is as follows:
\begin{equation}
q_t = 
\begin{cases}
q_{t-1} \times \gamma & \text{if } E_t \leq E_{\text{thresh}} \\
q_{t-1} / \gamma & \text{if } E_t > E_{\text{thresh}} \\
q_{t-1} & \text{if } E_{\text{thresh}} - \epsilon < E_t < E_{\text{thresh}} + \epsilon
\end{cases}
\end{equation}
where \( \epsilon \) is a small tolerance parameter to prevent oscillations near the threshold.

The quantization process for \( \Delta g_m^{(k)} \) introduces a quantization error, \( e_m^{(k)} \), defined as:
\begin{equation}
e_m^{(k)} = \Delta g_m^{(k)} - Q(\Delta g_m^{(k)})
\end{equation}
This error is bounded by the quantization level:
\begin{equation}
\|e_m^{(k)}\|_2 \leq \frac{R}{2^b}
\end{equation}
In our model, the dynamic adjustment of \( q_t \) influences the quantization granularity and, consequently, the reconstruction error \( e_m^{(k)} \). To detect when the model has reached convergence and reduce oscillations, the model monitors the average \( E_t \) over a window of \( W \) rounds. Define the moving average \( \overline{E}_t \) as:
\begin{equation}
\overline{E}_t = \frac{1}{W} \sum_{k=t-W+1}^{t} E_k
\end{equation}
If \( \overline{E}_t \leq E_{\text{thresh}} \) for \( W \) consecutive rounds, \( q_t \) is held constant to prevent unnecessary quantization adjustments. Periodically, proposed model re-evaluates the initial quantization level \( q_{\text{init}} \) by computing the variance \( V_t \) of \( E_t \) over the last \( T \) rounds. If \( V_t \) exceeds a threshold \( V_{\text{thresh}} \), the system resets \( q_{\text{init}} \) to a new baseline that reflects current conditions.
\begin{algorithm}
\caption{Adaptive Gradient Innovation and Error Sensitivity-Based Quantization with Communication Optimization}
\label{alg:alg1}
\begin{algorithmic}[1]
\State \textbf{Initialize:} Model parameters \( w^{(0)} \), quantization level \( q_t \), threshold \( \epsilon \), skip counter \( t_m = 0 \)
\For{each client \( m \) at iteration \( k \)}
    \State Compute local gradient: \( \nabla f_m^{(k)}(w) \)
    \State Calculate gradient innovation: \( \Delta g_m^{(k)} = \nabla f_m^{(k)}(w) - Q(\nabla f_m^{(k-1)}(w)) \)
    \State Quantize innovation: \( \Delta Q_m^{(k)} = Q(\Delta g_m^{(k)}) \)
    \State \textbf{Error Sensitivity Update:}
    \State Compute error sensitivity \( E_t = \frac{1}{|S_t|} \sum_{i \in S_t} \left\| \nabla F_i(\mathbf{w}_t) - G_t \right\|^2 \)
    \If{\( E_t > E_{\text{thresh}} \)} \State Update \( q_t = q_{t-1} / \gamma \)
    \Else \If{\( E_t \leq E_{\text{thresh}} \)} \State Update \( q_t = q_{t-1} \times \gamma \)
    \EndIf
    \EndIf
    \State \textbf{Communication Rule:}
    \If{\( \|\Delta Q_m^{(k)}\|_2 \geq \epsilon \) \textbf{or} \( t_m \geq \tau \)}
        \State Send update, reset \( t_m = 0 \)
    \Else
        \State Skip update, increment \( t_m = t_m + 1 \)
    \EndIf
    \State Server aggregates: \( G^{(k)} = \sum_{m \in K_{\text{active}}} \Delta Q_m^{(k)} \)
    \State Update model: \( w_{k+1} = w_k - \eta G^{(k)} \)
\EndFor
\end{algorithmic}
\end{algorithm}

\subsection{Communication Frequency Optmization}

The Communication Frequency optimization in our proposed framework with quantization strategy is designed to reduce unnecessary data transmission by allowing clients to selectively withhold updates that do not significantly contribute to the model's progress. Each client \( m \) computes a quantized gradient difference (innovation) \( \Delta Q_m^{(k)} \) at each iteration \( k \). Gradient difference threshold ensures that clients only transmit updates if their gradient innovations exceed a minimum level of significance, measured by the \( \ell_2 \)-norm of the quantized gradient difference. The gradient difference threshold \( \epsilon \) is defined such that:

\begin{equation}
\|\Delta Q_m^{(k)}\|_2 \geq \epsilon
\end{equation}

- \( \|\Delta Q_m^{(k)}\|_2 = \left( \sum_{i=1}^d (\Delta Q_{m,i}^{(k)})^2 \right)^{1/2} \)
- \( d \) is the dimensionality of the gradient vector, and \( \Delta Q_{m,i}^{(k)} \) is the \( i \)-th component of the quantized gradient difference at iteration \( k \) for client \( m \).

In this context, the gradient difference \( \Delta Q_m^{(k)} \) is defined as the change in the client’s quantized gradient compared to the previous iteration:
\begin{equation}
\Delta Q_m^{(k)} = Q(\nabla f^{(k)}_m(w_k)) - Q(\nabla f^{(k-1)}_m(w_{k-1}))
\end{equation}

Moreover, the Iteration Count Condition is our model prevent clients from skipping updates indefinitely, enforces a maximum skip period. Each client maintains a skip counter \( t_m \), which tracks the number of consecutive iterations since it last communicated an update to the server. When this count reaches a pre-defined maximum \( \tau \), the client is required to transmit its update, regardless of the gradient difference threshold. The communication rule for client \( m \) at iteration \( k \) can be defined as follows:
\begin{itemize}
    \item \[
        \|\Delta Q_m^{(k)}\|_2 < \epsilon, \quad t_m < \tau
        \quad \Rightarrow \quad \textit{skip update}, \ t_m \gets t_m + 1
    \]
    \item \[
        \|\Delta Q_m^{(k)}\|_2 \geq \epsilon, \quad t_m \geq \tau
        \quad \Rightarrow \quad \textit{send update}, \ t_m \gets 0
    \]
\end{itemize}

The server aggregates only significant gradient innovations from clients, conserving bandwidth and computation. A threshold \( \epsilon \) is applied to the \( \ell_2 \)-norm of quantized gradient innovations. If a client \( m \)’s update at iteration \( k \) meets:
\begin{equation}
\|\Delta Q_m^{(k)}\|_2 \geq \epsilon
\end{equation}

where \( \Delta Q_m^{(k)} = Q(\Delta g_m^{(k)}) \), it is included in the aggregation. This ensures minor updates do not affect the global gradient, reducing computational load. The global gradient \( G^{(k)} \) is computed as:
\begin{equation}
G^{(k)} = \sum_{m \in K_{\text{active}}} \Delta Q_m^{(k)}
\end{equation}

where \( K_{\text{active}} \) includes clients meeting the threshold. The server updates model parameters with learning rate \( \eta \):
\begin{equation}
w_{k+1} = w_k - \eta G^{(k)}
\end{equation}

This selective aggregation ensures each update is meaningful, conserving resources while maintaining model performance. The gradient quantization and communication optimization process is presented in Algorithm\textcolor{blue}{~\ref{alg:alg1}}.
\begin{figure*}[!ht]
\centering
\includegraphics[width=15.8cm,height=13cm,keepaspectratio]{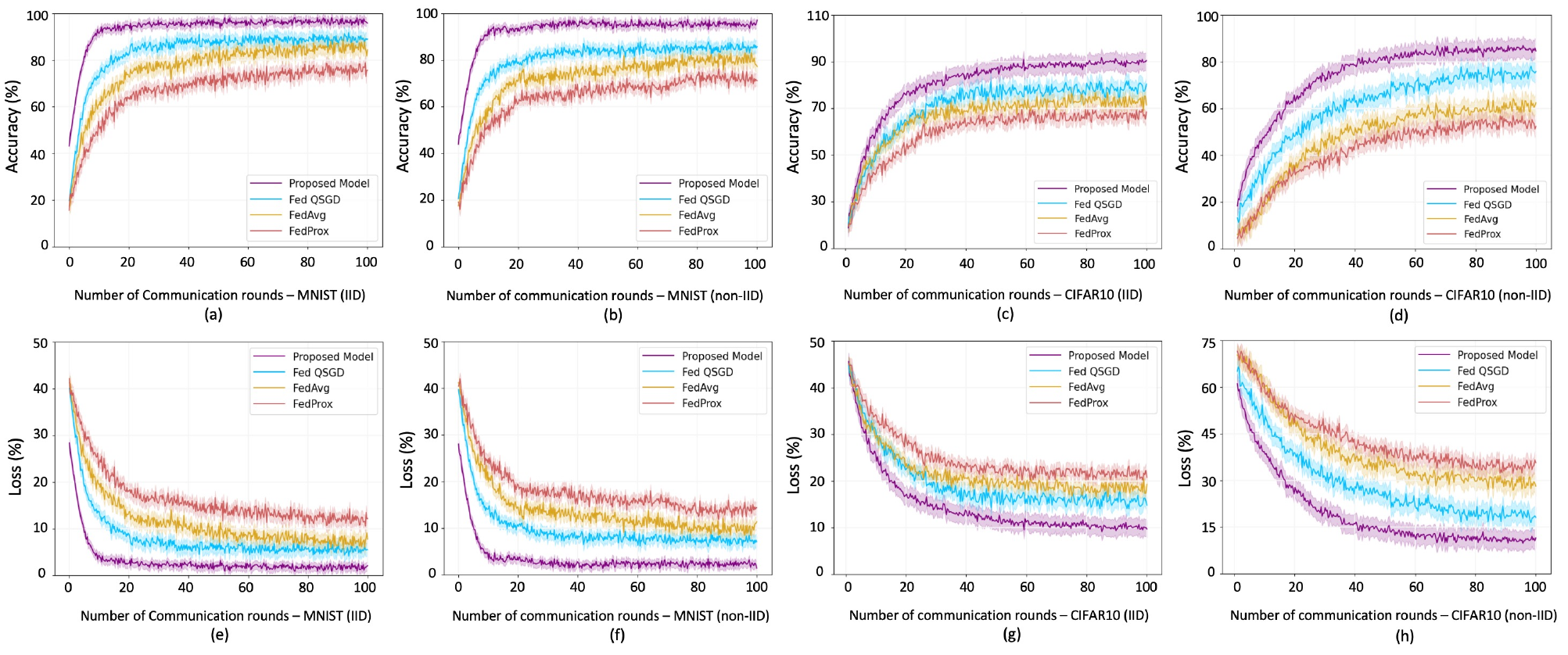}
\caption{Performance analysis of accuracy, loss, and convergence for the proposed and comparative schemes on both MNIST and CIFAR10 datasets under IID and non-IID data settings.}
\label{fig2}
\end{figure*}

\begin{figure}[!ht]
\centering
\includegraphics[width=8.2cm,height=10cm,keepaspectratio]{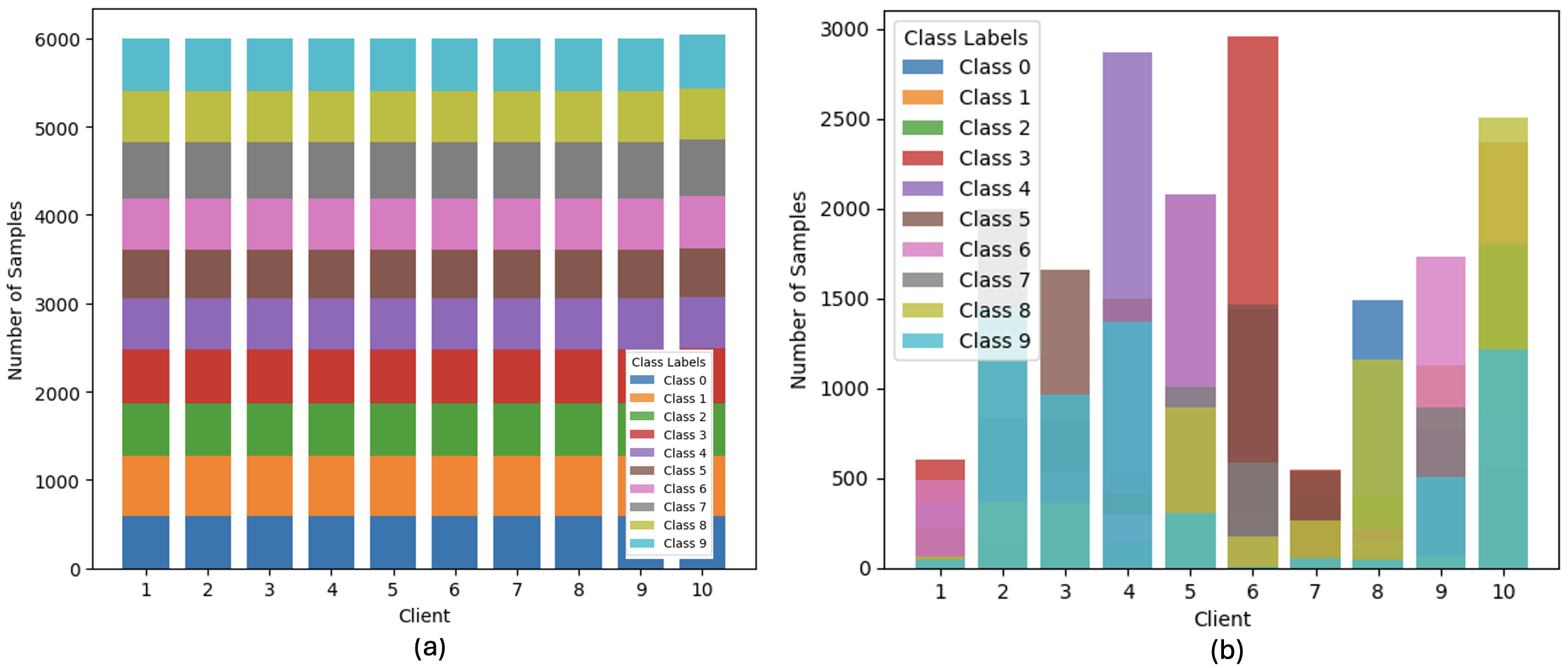}
\caption{The data distribution used for comparison analysis, (a) Startified sampling (IID), (b) Dirichlet sampling (non-IID)}
\label{fig3}
\end{figure}

\begin{figure}[!ht]
\centering
\includegraphics[width=8.2cm,height=10cm,keepaspectratio]{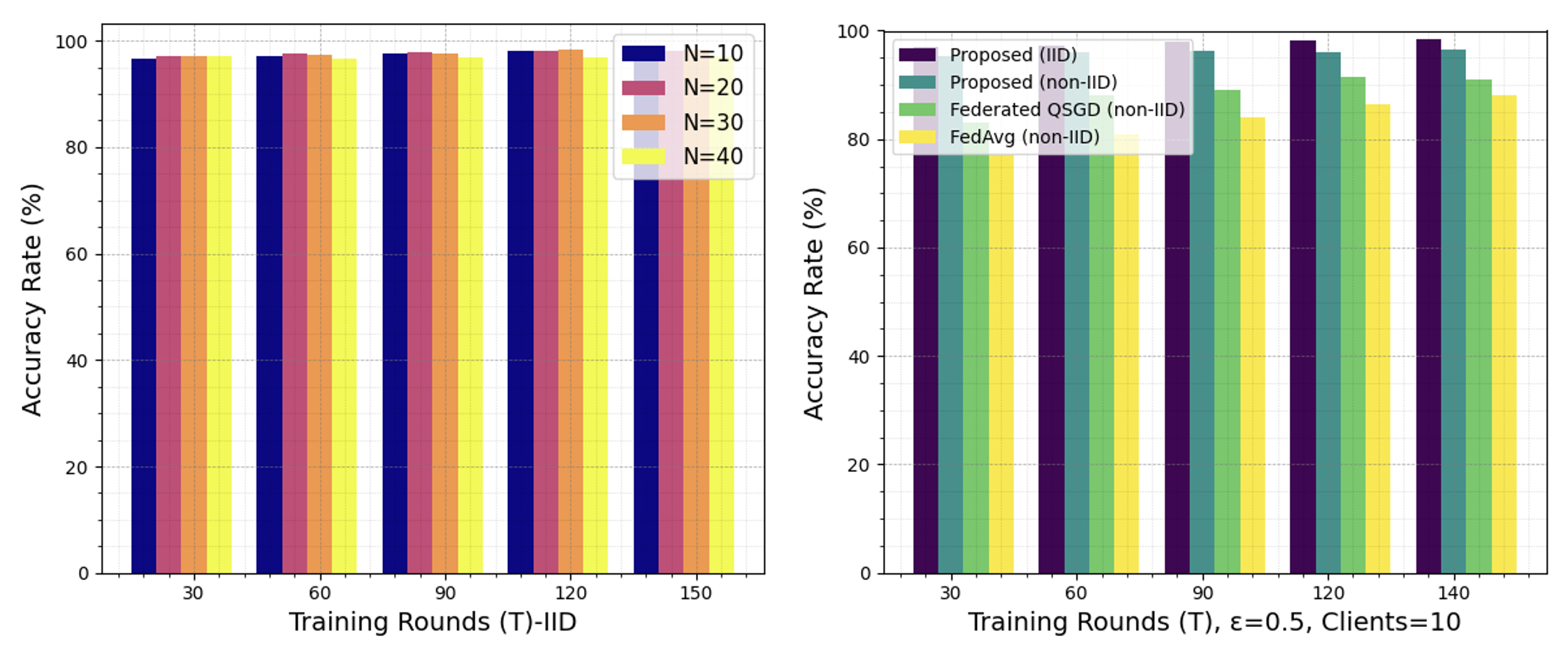}
\caption{ (a) Accuracy performance analysis of proposed framework with different number of clients, (b) Accuracy Performance Analysis of the Proposed Framework (IID and Non-IID) with Comparative Schemes under IID Data Distribution.}
\label{fig4}
\end{figure}

\section{Results}
In this section we will present the performance analysis of our proposed framework and compare it with comparative schemes. We evaluated the performance of our proposed framework using accuracy and loss metrics, and also present a convergence analysis. The framework is assessed with performance metrics such as the number of clients, communication rounds, and various data distributions. We conducted evaluations on both Independent and Identically Distributed (IID) based on stratified sampling and non-IID data distributions using a Dirichlet distribution to analyze performance across different data settings. The illustration of data sampling is presented in Fig.\textcolor{blue}{~\ref{fig3}}.

The datasets used for evaluation is the MNIST and CIFAR-10. MNIST is a handwritten digit dataset (28x28 images) with 70,000 images across 10 classes, divided into 60,000 for training and 10,000 for testing. CIFAR10 is image classification dataset consisting of 60,000 32x32 color images in 10 different classes, with 6,000 images per class. The model chosen for evaluation is a Convolutional Neural Network (CNN) for MNIST and Resnet-20 for CIFAR-10. In the generalized experiment, we used 10 clients, but we varied the number of clients to obtain more diverse results. Moreover, we adopted Adam optimizer with a learning rate of 0.05.

In Fig.\textcolor{blue}{~\ref{fig2}}, we presented the accuracy and loss performance analysis of the proposed framework over both IID and non-IID settings. Moreover, we compared the effectiveness of our framework with baseline models, Federated QSGD and Federated Averaging (FedAvg) and fedProx. Fig.\textcolor{blue}{~\ref{fig2}}(a) and Fig.\textcolor{blue}{~\ref{fig2}}(c) evaluate the accuarcy performance of our proposed work with IID data distribution, while Fig.\textcolor{blue}{~\ref{fig2}}(b) and Fig.\textcolor{blue}{~\ref{fig2}}(d) evaluate the performance with non-IID data distribution across a number of communication rounds for both MNIST and CIFAR-10 datasets. The total number of clients is 10. The results show that the proposed framework in this paper outperforms the comparative schemes. Moreover, it is clearly shown in the results that the proposed framework is converging faster and taking fewer communication rounds to achieve convergence. The Adaptive Feature-Elimination Strategy drops less informative vectors to improve efficiency. Because of the efficient Adaptive Gradient Innovation and Error Sensitivity-Based Quantization and Communication Frequency Optimization, significant contributions are made to model improvements. Fig.\textcolor{blue}{~\ref{fig2}}(e) and Fig.\textcolor{blue}{~\ref{fig2}}(g) evaluate the loss performance of our proposed work with IID data distribution, while Fig.\textcolor{blue}{~\ref{fig2}}(f) and Fig.\textcolor{blue}{~\ref{fig2}}(h) evaluate the performance with non-IID data distribution across a number of communication rounds for both MNIST and CIFAR-10 datasets. The results show that the loss performance of our proposed framework is also outperforming other comparative schemes. The Federated QSGD performs better than FedAvg and fedProx because of its time-adaptive quantization. Moreover, the model is taking fewer communication rounds to achieve convergence, resulting in less communication overhead. The performance results show that the accuracy difference of the proposed framework with IID and non-IID data distribution is not very significant. However, the comparative schemes show a greater difference in accuracy and loss performance when comparing IID and non-IID data distributions. Additionally, comparative schemes are taking longer communication rounds for convergence.

Fig.\textcolor{blue}{~\ref{fig4}}(a) evaluates the accuracy performance of the proposed framework over different numbers of clients, where the number of clients is set to 10, 20, 30, and 40 to assess dense network performance. The graph shows that increasing the number of clients does not significantly affect the performance results. The results indicate a slight increase in accuracy with the increase in the number of clients. In Fig.\textcolor{blue}{~\ref{fig2}}(b), we evaluated the accuracy performance of the proposed framework (both IID and non-IID), Fed QSGD (IID), and FedAvg (IID). The results show that due to the efficient Adaptive Gradient Innovation and Error Sensitivity-Based Quantization and Communication Frequency Optimization, there are significant improvements in the proposed framework; even non-IID data results are better compared to the comparative schemes. We did not evaluate the performance with diverse datasets to check the effectiveness of our proposed model. Moreover, a limitation remains, as accuracy with non-IID datasets could be further improved. We observed that a higher number of aggregations for imbalanced datasets results in improved accuracy. In our future work, we will investigate these points to further enhance the quality of the framework for a more communication-efficient framework that balances accuracy and loss performance as well. 

\section{Conclusion}

In this paper, we propose a three-fold communication-efficient strategy for FL, including an Adaptive Feature-Elimination Strategy, an Adaptive Gradient Innovation Quantization, which compresses gradient changes and adjusts quantization levels based on error sensitivity, and a Gradient Communication Frequency Optimization strategy. In our paper, we improved the performance of our proposed model in terms of communication efficiency, achieving high accuracy with IID datasets. However, there remains a limitation, as accuracy with non-IID datasets could be further improved. We observed that a higher number of aggregations for imbalanced datasets results in improved accuracy. To address this, we plan to extend this work by incorporating the HFL concept alongside Edge FL to bring computation and aggregation closer to the node, reducing the computational expense and communication overhead associated with frequent gradient exchanges with the server.

\begin{IEEEbiography}[{\includegraphics[width=1in,height=1.25in,clip,keepaspectratio]{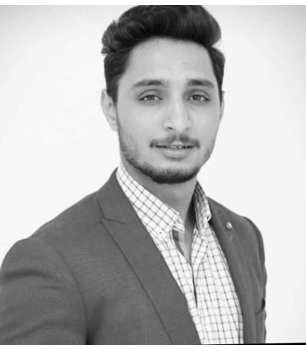}}]{Asadullah Tariq (M'23)} completed his Bachelor's and Master's degrees in Computer Science from PUCIT and FAST-NUCES, Pakistan, in 2016 and 2019, respectively. He is currently pursuing his Ph.D. in Computer Science at the United Arab Emirates University, UAE. He has authored numerous journal papers, conference papers and book chapters with notable contributions to prestigious journals such as \textit{IEEE COMST} (IF: 35.6), \textit{IEEE OJCOMS}, \textit{IEEE IoT J., IEEE TCE, IEEE Access, Applied Sciences} and top conferences such as \textit{IEEE INFOCOM, IEEE ICC, ACM/IEEE SEC, IEEE IWCMC, IEEE FNWF and EAI SecureComm etc.}. He got best paper award at \textit{IEEE IIT’23} and \textit{EAI SecureComm'24}, Student travel grant at \textit{ACM/IEEE SEC'23} and best presentation award at \textit{BDSIC'23}. He is the author of two published book chapters in Elsevier. He has held an adjunct faculty position at UAE University since 2023. His research interests include, but are not limited to, Federated Learning, Machine Learning, Wireless Communication, Edge AI, NDN-ICN, LLM and IoT. Currently, his research focus is on developing frameworks for trustworthiness in federated learning. Furthermore, Asadullah plays an active role as a reviewer for reputed journals, including but not limited to \textit{IEEE COMST, IEEE Commun. Mag., IEEE Comm. Letters, IEEE TAI, IEEE JBHI, IEEE TCC, IEEE TITS, JNCA, IEEE IoT J., IEEE Access, Elsevier Information Sciences, FGCS} and others.
\end{IEEEbiography}
\begin{IEEEbiography}[{\includegraphics[width=1in,height=1.25in,clip,keepaspectratio]{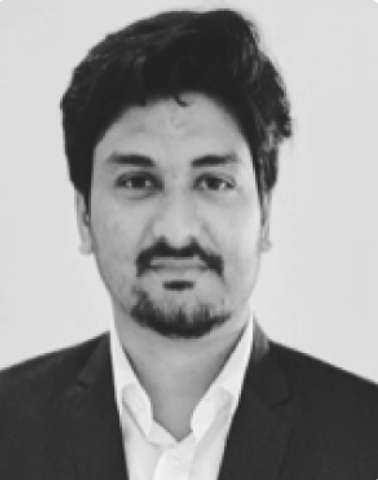}}]{Tariq Qayyum (M'23)}\, is currently pursuing the Ph.D. degree with the College of Information Technology, United Arab Emirates University (UAEU). His research interests include federated learning, network security, cloud computing, fog computing, the Internet of Things (IoT), and distributed systems. He has authored numerous journal papers, conference papers and book chapters with notable contributions to prestigious journals such as \textit{IEEE IoT J., IEEE Trans. on Sustain. Comp., IEEE Access, IEEE OJCOMS, SN Cloud Computing} and top conferences such as \textit{IEEE INFOCOM, IEEE ICC, ACM/IEEE SEC, IEEE IWCMC, IEEE FNWF and EAI SecureComm etc.}.
\end{IEEEbiography}
\begin{IEEEbiography}[{\includegraphics[width=1in,height=1.25in,clip,keepaspectratio]{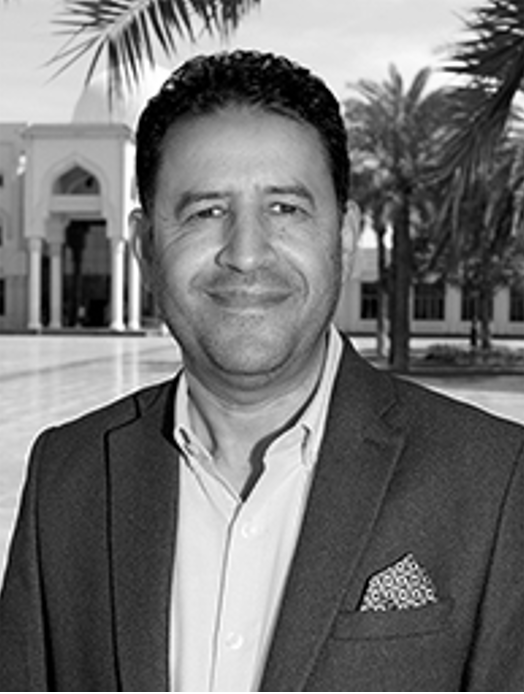}}]{M.A.Serhani} received the Ph.D. degree in computer engineering from Concordia University, in 2006, the M.Sc. degree in software engineering from the University of Montreal, Canada, in 2002, and the B.Sc. in Computer Science from ESCI, Morocco, in 1996. He is currently a Professor with the College of Computing and Informatics, University of Sharjah. Before joining UoS, Dr. Serhani was an Assistant Dean for Research and Graduate Studies at the College of Information Technology, UAE University between 2018 and 2022, a Ph.D. program director from 2018 to 2020, and a Master program director from 2014 to 2017. His research interests include Federated Learning, Cloud for data intensive e-health applications, and services; SLA enforcement in Cloud data centers, and big data value chain; Cloud federation and monitoring, and non-invasive smart health monitoring; management of communities of web services; and web services applications and security. Dr. Serhani has a large experience earned throughout his involvement and management of different research and development projects. He has served as organizer and technical program chair/co-chair of many international conferences, and workshops (e.g., IWCMC’23, BDSIC’23, SERVICES’21, ICWS’20, AI-PA’2020, IIT’13, NDT’12, WMCS’12). He has published around 170 refereed publications, including journals, conferences, a book, and book chapters. He was granted several awards and recognitions including the CIT merit allowance for four times, CIT excellence award in research in 2015, CIT excellence award in teaching in 2016, the CIT excellence award in service in 2019.
\end{IEEEbiography}

\begin{IEEEbiography}[{\includegraphics[width=1in,height=1.25in,clip,keepaspectratio]{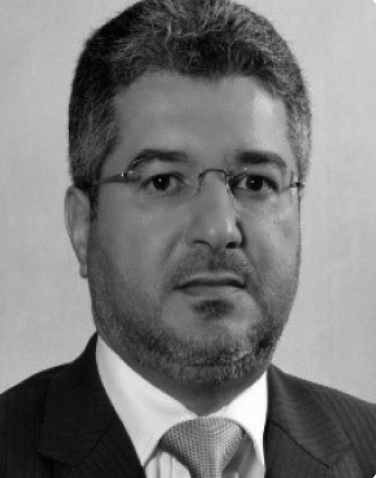}}]{Farag Sallabi} received the Ph.D. degree in electrical and computer engineering from the University of Ottawa, Canada, in 2001. He is an Associate Professor with the Department of Computer and Network Engineering and the Coordinator of the Ph.D. Program of Informatics and Computing with the CIT, United Arab Emirates University (UAEU). He was with Sigpro Wireless Company, Ottawa, Canada, from August 2001 to August 2002. He was with UAEU, from September 2002 to June 2012. From July 2012 to July 2015, he was the Director General of the Directorate General of Electronic Services, Ministry of Communications and Informatics, Libya. He has led a project to develop and implement an e-Government program in Libya. He then rejoined the UAEU, in August 2015. His research interests include quality of service provisioning in wired and wireless networks, routing in wireless sensor networks, performance evaluation of WSN, network management, modeling and simulation, the IoT, and smart healthcare systems. 
\end{IEEEbiography}

\begin{IEEEbiography}[{\includegraphics[width=1in,height=1.25in,clip,keepaspectratio]{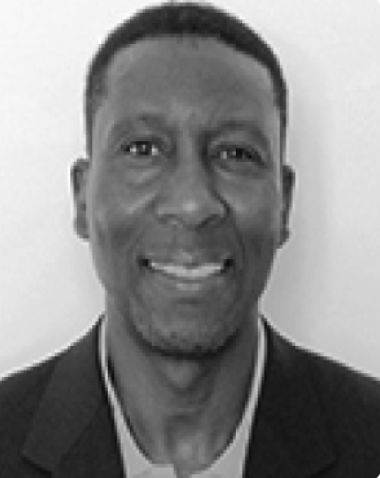}}]{Ezedin S. barka}\,received the Ph.D. degree in information technology from George Mason University, Fairfax, VA, USA, in 2002. He is currently an Associate Professor and the Department Chair with the College of Information Technology, United Arab Emirates University. He was a member of the Laboratory for Information Security Technology (LIST), George Mason University. He has published over 50 journals and conference papers. His current research interests include access control, where he published a number of papers addressing delegation of rights using RBAC. Other research interests include digital rights management (DRM), large-scale security architectures and models, trust management, security in UAVs, and network wired and wireless and distributed systems security. He is a member of the IEEE Communications Society and the IEEE Communications and Information Security Technical Committee (CISTC). He serves on the technical program committees for many international IEEE conferences, such as ACSAC, GLOBECOM, ICC, WIMOB, and WCNC. In addition, he was a reviewer of several international journals and conferences.
\end{IEEEbiography}
\begin{IEEEbiography}[{\includegraphics[width=1in,height=1.25in,clip,keepaspectratio]{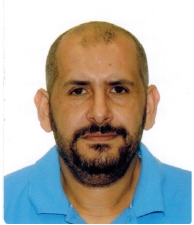}}]{Ikbal Taleb} holds a Ph.D. in Information and Systems Engineering from Concordia University, and MSc. in Software Engineering from University of Montreal, Canada. His research interests include Big data quality, quality assessment, quality profiling, and web services. He has more than twenty years in Industry and Academia. Presently, he works with the College of Technological Innovations (CTI) at Zayed University in Abu Dhabi.
\end{IEEEbiography}


\begin{thebibliography}{00}
\bibitem{b1} 
Tariq, A., Serhani, M.A., Sallabi, F.M., Barka, E.S., Qayyum, T., Khater, H.M. and Shuaib, K.A., Trustworthy Federated Learning: A Comprehensive Review, Architecture, Key Challenges, and Future Research Prospects. IEEE Open Journal of the Communications Society, vol. 5, pp. 4920-4998, 2024.
\bibitem{b2}
 Zhang, X., Zhu, X., Wang, J., Yan, H., Chen, H. and Bao, W., Federated learning with adaptive communication compression under dynamic bandwidth and unreliable networks. Information Sciences, 540, pp.242-262, 2020.
\bibitem{b3}
Almanifi, O.R.A., Chow, C.O., Tham, M.L., Chuah, J.H. and Kanesan, J., Communication and computation efficiency in federated learning: A survey. Internet of Things, 22, p.100742, 2023.
\bibitem{b4}
Shah, S.M. and Lau, V.K., Model compression for communication efficient federated learning. IEEE Transactions on Neural Networks and Learning Systems, 34(9), pp.5937-5951, 2021.
\bibitem{b5}
Lan, M., Ling, Q., Xiao, S. and Zhang, W., Quantization bits allocation for wireless federated learning. IEEE Transactions on Wireless Communications, 22(11), pp.8336-8351, 2023.
\bibitem{b6} 
Jiang, X. and Borcea, C., Complement sparsification: Low-overhead model pruning for federated learning. In Proceedings of the AAAI Conference on Artificial Intelligence (Vol. 37, No. 7, pp. 8087-8095), 2023.

\bibitem{gq1}
Reisizadeh, A., Mokhtari, A., Hassani, H., Jadbabaie, A. and Pedarsani, R., Fedpaq: A communication-efficient federated learning method with periodic averaging and quantization. In International conference on artificial intelligence and statistics (pp. 2021-2031). PMLR, 2020.

\bibitem{gq2}
Mao, Y., Zhao, Z., Yan, G., Liu, Y., Lan, T., Song, L. and Ding, W., Communication-efficient federated learning with adaptive quantization. ACM Transactions on Intelligent Systems and Technology (TIST), 13(4), pp.1-26, 2022.

\bibitem{gq3}
Hönig, R., Zhao, Y. and Mullins, R., DAdaQuant: Doubly-adaptive quantization for communication-efficient federated learning. In International Conference on Machine Learning (pp. 8852-8866). PMLR, 2022.

\bibitem{gq4}
Oh, Y., Lee, N., Jeon, Y.S. and Poor, H.V., Communication-efficient federated learning via quantized compressed sensing. IEEE Transactions on Wireless Communications, 22(2), pp.1087-1100, 2022.
\bibitem{gq5}
J. Sun, T. Chen, G. B. Giannakis, Q. Yang and Z. Yang, "Lazily Aggregated Quantized Gradient Innovation for Communication-Efficient Federated Learning," in IEEE Transactions on Pattern Analysis and Machine Intelligence, vol. 44, no. 4, pp. 2031-2044, 2022.
\bibitem{gq6}
Liu, W., Chen, L. and Zhang, W.,. Decentralized federated learning: Balancing communication and computing costs. IEEE Transactions on Signal and Information Processing over Networks, 8, pp.131-143, 2022.

\bibitem{mq4}
Qu, L., Song, S. and Tsui, C.Y., 2022, December. FedDQ: Communication-efficient federated learning with descending quantization. In GLOBECOM 2022 - 2022 IEEE Global Communications Conference, Rio de Janeiro, Brazil, 2022, pp. 281-286, 2022. doi: 10.1109/GLOBECOM48099.2022.10001205.

\bibitem{a1}
A. Tariq, F. Sallabi, M. A. Serhani, T. Qayyum and E. S. Barka, "Leveraging Game Theory and XAI for Data Quality-Driven Sample and Client Selection in Trustworthy Split Federated Learning," in IEEE Transactions on Consumer Electronics, vol. 71, no. 2, pp. 6686-6699, May 2025, doi: 10.1109/TCE.2025.3543209.
\bibitem{a2}
M. A. Serhani, H. G. Abreha, A. Tariq, M. Hayajneh, Y. Xu and K. Hayawi, "Dynamic Data Sample Selection and Scheduling in Edge Federated Learning," in IEEE Open Journal of the Communications Society, vol. 4, pp. 2133-2149, 2023, doi: 10.1109/OJCOMS.2023.3313257.

\bibitem{gq8}
Liu, H., He, F. and Cao, G., Communication-efficient federated learning for heterogeneous edge devices based on adaptive gradient quantization. IEEE INFOCOM 2023 - IEEE Conference on Computer Communications, New York City, NY, USA, 2023, pp. 1-10, doi: 10.1109/INFOCOM53939.2023.10228970.
\bibitem{new}
Tariq, A., Rehman, R.A. and Kim, B.S., 2021. An Intelligent Forwarding Strategy in SDN-Enabled Named-Data IoV. Computers, Materials \& Continua, 69(3).

\bibitem{gq9}
Oh, J., Lee, D., Won, D., Noh, W. and Cho, S., Communication-efficient federated learning over-the-air with sparse one-bit quantization. IEEE Transactions on Wireless Communications, vol. 23, no. 10, pp. 15673-15689, 2024. 
\end{thebibliography}
\end{document}